\documentclass[showpacs,aps,showkeys,graphicx,twocolumn,]{revtex4}
\usepackage{amsmath}
\usepackage{mathrsfs}
\usepackage{amsfonts}
\usepackage{amssymb}
\usepackage{graphicx}
\usepackage{eufrak}
\usepackage{multirow}
\usepackage{longtable}
\usepackage{float}
\usepackage{bm}
\usepackage{soul,color,xcolor}

\begin{document}

\title{Collective unitary evolution with linear optics by Cartan decomposition}

\author{Wen-Qiang Liu\textsuperscript{1,2}, Xin-Jie Zhou\textsuperscript{1}, and Hai-Rui Wei\textsuperscript{1}\footnote{Corresponding author: hrwei@ustb.edu.cn}}

\address{1 School of Mathematics and Physics, University of Science and Technology Beijing, Beijing 100083, China \\
2 Center for Quantum Technology Research and Key Laboratory of Advanced Optoelectronic Quantum Architecture and Measurements (MOE), School of Physics, Beijing Institute of Technology, Beijing 100081, China}

\date{\today }

\begin{abstract}
Unitary operation is an essential step for quantum information processing. We first propose an iterative procedure for decomposing a general unitary operation without resorting to controlled-NOT gate and single-qubit rotation library.  Based on the results of decomposition, we design two compact architectures to deterministically implement arbitrary two-qubit polarization-spatial and spatial-polarization collective unitary operations, respectively. The involved linear optical elements are reduced from 25 to 20 and 21 to 20, respectively. Moreover, the parameterized quantum computation can be flexibly manipulated by wave plates and phase shifters. As an application, we construct the specific quantum circuits to realize two-dimensional quantum walk and quantum Fourier transformation.  Our schemes are simple and feasible with the current technology.
\end{abstract}


\maketitle

\section{Introduction}   \label{sec1}

Linear optics provides an alternative physical platform for realizing varied quantum technologies, ranging from quantum computing \cite{KLM,LOQC,Liu}, quantum communication \cite{communication1,QSDC1,Teleportation3,QSDC0,QSDC,QSDC2,EPP14,qkd,QSDC3,QSDC4,QSDC5} to quantum algorithms \cite{Shor1,Equations} and quantum metrology \cite{Metrology,Metrology1}. Photon is considered as an outstanding candidate for complex quantum information processing (QIP) because it naturally possesses various inherent qubit-like degrees of freedom (DOFs), long coherence time, robustness against the decoherence, and the availability of manipulation at single-photon level  \cite{manipulate1,LOQC}.
Small-scale quantum computing with linear optics can be implemented probabilistically \cite{KLM}. To date, ambitions to realize multi-qubit photonic quantum computing on only one DOF are still obstructed since the physical interaction between individual photons goes beyond those currently available.

QIP with multiple DOFs of photons is a promising research field for implementing effectively large-scale quantum computation and quantum communication, due to its appealing features, such as deterministic property \cite{Deterministic,Deterministic1}, high capacity of quantum channel \cite{capacity1}, low photon loss rate \cite{Low-loss}, reduced quantum resources \cite{Lanyon}, and flexible interactions between different DOF qubits \cite{P-M-Cluster1,P-M-Cluster2,P-M-Cluster3,hyperentanglement}.  Multiple DOFs of a single photon have been widely used in linear optical QIP. For instance, the implementation of spatial-polarization Deutsch--Jozsa algorithm have been reported by Scholz \emph{et al}. \cite{DJ1} in 2006 and Zhang \emph{et al}. \cite{DJ2} in 2010. The creation of four-qubit and six-qubit polarization-momentum entangled cluster states have been experimentally demonstrated in recent years \cite{P-M-Cluster1,P-M-Cluster2,P-M-Cluster3}.
In 2010, Sheng \emph{et al}. \cite{Sheng} first proposed hyperentanglement Bell-state analysis and quantum teleportation in the polarization and spatial DOFs. Subsequently, Wang \emph{et al}. \cite{teleportation-2} experimentally demonstrated the quantum teleportation in both polarization and orbital angular momentum DOFs.
In 2012, Abouraddy \emph{et al}. \cite{orbital2} proposed universal polarization-momentum optical entangled quantum gates and the generation of entangled states. Later, a single-photon time-polarization controlled-phase gate was also experimentally demonstrated \cite{CPF-time-polarization}.   In 2015, based on cosine-sine decomposition (CSD) \cite{CSD}, Goyal \emph{et al}. \cite{quantum-walk} proposed an interesting scheme to complete spatial-polarization quantum walk in single photon level. Subsequently, the CSD algorithm  was applied to realize arbitrary discrete unitary spatial-polarization transformation by using linear optics \cite{Dhand-Goyal}. In 2019, arbitrary spatial-temporal collective unitary operation was theoretically reported \cite{Su,Kumar}.  In 2021,  deterministic polarization-orbital-angular-momentum Toffoli gate and Fredkin gate with a single photon have been experimentally demonstrated \cite{hybrid-Toffoli,hybrid-Fredkin}.
We also note that multiple DOFs of photons have been widely used in entanglement purification \cite{Deterministic,Deterministic1} and concentration \cite{concentration}, quantum secure direct communication (QSDC) \cite{QSDC6,QSDC7}, quantum key distribution (QKD) \cite{qKD}, and etc.

Adding optical elements to the quantum circuit increases its overall imperfections, which prevents the realization of quantum computation within sufficient precision and also poses a challenge for the stability of the circuits. It is therefore crucial to construct  efficient quantum circuit with significantly fewer linear optics.
In this letter, we aim to reduce the cost of the linear-optics-based arbitrary collective unitary operation using polarization and spatial-mode DOFs of a photon. We first reconstruct a universal unitary operation $U \in$ SU$(2^n)$ by Cartan decomposition technique. Based on the results of decomposition, we realize arbitrary two-qubit linear optical collective unitary operation with the polarization-spatial DOF and spatial-polarization DOF of a single photon, respectively.  As an application, we design the specific schemes for implementing two-dimensional quantum walk and quantum Fourier transformation.

Our schemes have the following characteristics: (i) Our schemes are to reduce the number of required linear optical elements and are not constructed in terms of CNOT gates and single-qubit rotations. (ii) The parameterized quantum computation can be easily manipulated by quarter-wave plates (QWPs), half-wave plates (HWPs) and phase shifters (PSs). (iii) The number of required linear optical elements for implementing two-qubit polarization-spatial  (spatial-polarization) collective unitary operation is reduced from 25 (21) to 20 (20). (iv) Our schemes reduce the quantum resource cost, are more robustness against the photonic dissipation, and feasible with the current technology.

\section{Arbitrary polarization-spatial collective unitary operation}\label{Sec2}

Cartan decomposition \cite{Cartan,Cartan1} derives from the Lie group and relies on its Lie algebra, and it is a decomposition of the Lie group of unitary evolutions. A Cartan decomposition of semi-simple real Lie algebra $\mathfrak{g}$ is a vector space decomposition
\begin{eqnarray}    \label{eq1}
\mathfrak{g}=\mathfrak{l}\oplus \mathfrak{p},
\end{eqnarray}
where Lie algebras $\mathfrak{l}$ and $\mathfrak{p}=\mathfrak{l}^\bot$ satisfy,
\begin{eqnarray}    \label{eq2}
[\mathfrak{l},\mathfrak{l}]\subseteq \mathfrak{l},\;\;\;
[\mathfrak{l},\mathfrak{p}]\subseteq \mathfrak{p},\;\;\;
[\mathfrak{p},\mathfrak{p}]\subseteq \mathfrak{l}.
\end{eqnarray}
Here the Lie bracket for matrix algebras $[\mathfrak{a},\mathfrak{b}]=\mathfrak{a}\mathfrak{b}-\mathfrak{b}\mathfrak{a}$.

According to the relationship between the Lie group $G$ and the Lie algebra $\mathfrak{g}$, any given unitary transformation $U\in G$ (element of $G=e^\mathfrak{g}$) can be decomposed as
\begin{eqnarray}    \label{eq3}
U=K_1\;A\;K_2.
\end{eqnarray}
Here $K_1, K_2\in e^\mathfrak{l}$ and $A\in e^\mathfrak{h}\subset e^\mathfrak{p}$. $\mathfrak{h}$ is called Cartan subalgebra contained in $\mathfrak{p}$, which is a maximal Abelian subalgebra.

Cartan decomposition is the most popular and powerful tool for constructing efficient quantum circuit due to its flexible $\mathfrak{l}$ and $\mathfrak{h}$. Such excellent property induces some particular decompositions of $\text{SU}(2^n)$ group for every $n$-qubit unitary evolution, such as Khaneja and Glaser decomposition (KGD) \cite{Cartan1}, concurrence canonical decomposition (CCD) \cite{CCD}, odd-even decomposition (OED) \cite{OED}, and quantum Shannon decomposition (QSD) \cite{QSD}. The best result is the QSD-based construction \cite{Dhand-Goyal}, which beats the QR-based one \cite{QR}.
In the following, we simplify arbitrary polarization-spatial collective unitary operation by employing new $\mathfrak{l}$ and  $\mathfrak{h}$.

For $n=1$, the basis of the Lie algebra $\mathfrak{su}(2)$ can be spanned by
\begin{eqnarray}    \label{eq4}
\begin{split}
\mathfrak{l}_{2^1}:=\text{span}\{\sigma_z\}, \quad
\mathfrak{p}_{2^1}:=\text{span}\{\sigma_x, \sigma_y \}.
\end{split}
\end{eqnarray}
The Cartan subalgebra $\mathfrak{h}_{2^1}$ is selected as
\begin{eqnarray}    \label{eq5}
\begin{split}
\mathfrak{h}_{2^1}&:=\text{span}\{\sigma_x\}.
\end{split}
\end{eqnarray}
Here $\sigma_x$, $\sigma_y$, and $\sigma_z$ represent the standard Pauli spin matrices.

For $n=2$, $\mathfrak{su}(4)$ can be divided into the following form
\begin{eqnarray}    \label{eq6}
\begin{split}
\mathfrak{l}_{2^2}&:=\text{span}\{\mathfrak{su}(2)\otimes\mathbb{I}_2 , \mathfrak{u}(2)\otimes\sigma_z \},\\
\mathfrak{p}_{2^2}&:=\text{span}\{\mathfrak{u}(2)\otimes\sigma_x, \mathfrak{u}(2)\otimes\sigma_y \}.
\end{split}
\end{eqnarray}
We select Cartan subalgebra $\mathfrak{h}_{2^2}$ as
\begin{eqnarray}    \label{eq7}
\begin{split}
\mathfrak{h}_{2^2}&:=\text{span}\{\mathbb{I}_2\otimes \mathfrak{h}_{2^1}, \sigma_z\otimes \mathfrak{h}_{2^1} \}.
\end{split}
\end{eqnarray}
Here $\mathfrak{u}(2):= \text{span} \{\mathfrak{su}(2), \mathbb{I}_2\}$,  $\mathbb{I}_2$ is a $2\times2$ identity matrix. 

\begin{figure}      [htb]
\begin{center}
\includegraphics[width=8.2 cm,angle=0]{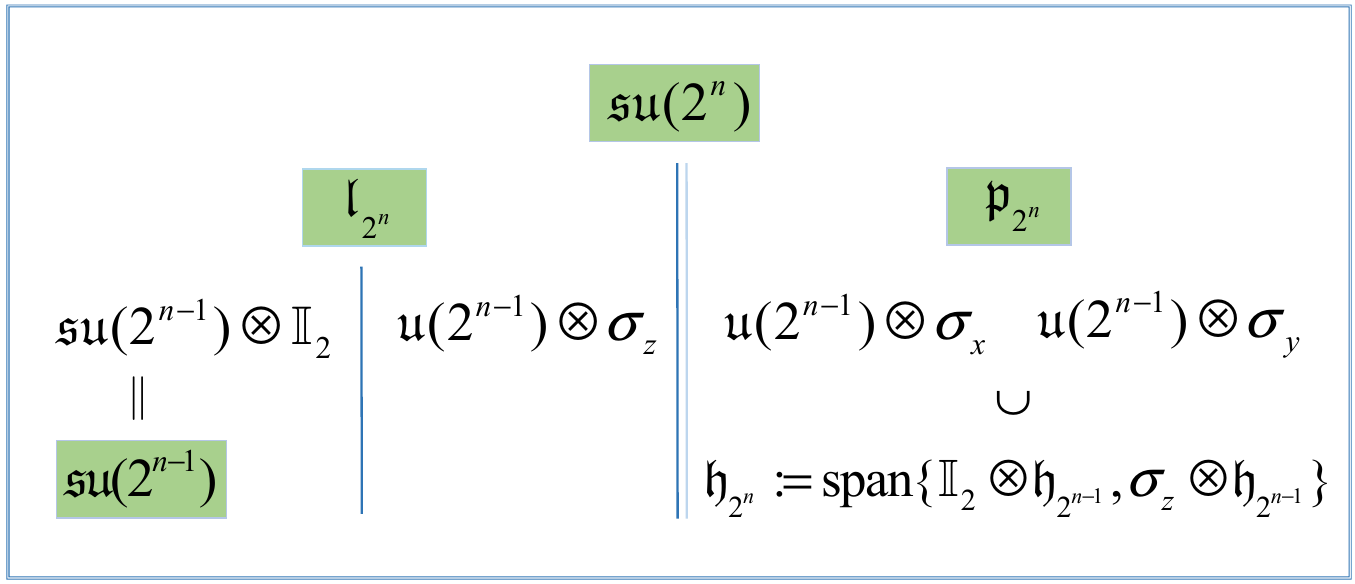}
\caption{ Schematic explanation for Cartan decomposition of a polarization-spatial collective unitary operation by a recursive algorithm. }  \label{su(2n)}
\end{center}
\end{figure}

For an $n$-qubit ($n\geq2$) case, as shown in  Fig. \ref{su(2n)},  the $n$th recurrence step  yields the $\mathfrak{su}(2^n)$ Cartan decomposition
\begin{eqnarray}    \label{eq9}
\begin{split}
\mathfrak{l}_{2^n}&:=\text{span}\{\mathfrak{su}(2^{n-1})\otimes\mathbb{I}_2, \mathfrak{u}(2^{n-1})\otimes\sigma_z\},\\
\mathfrak{p}_{2^n}&:=\text{span}\{\mathfrak{u}(2^{n-1})\otimes \sigma_x, \mathfrak{u}(2^{n-1})\otimes \sigma_y\}.
\end{split}
\end{eqnarray}
Cartan subalgebra $\mathfrak{h}_{2^n}$ can be taken as
\begin{eqnarray}    \label{eq10}
\begin{split}
\mathfrak{h}_{2^n}&:=\text{span}\{\mathbb{I}_2 \otimes \mathfrak{h}_{2^{n-1}},  \sigma_z\otimes \mathfrak{h}_{2^{n-1}} \}.
\end{split}
\end{eqnarray}
Here $\mathfrak{u}(2^{n-1}):= \text{span} \{\mathfrak{su}(2^{n-1}), \mathbb{I}_2^{\otimes{(n-1)}}\}$.

\subsection{Implementation of arbitrary two-qubit polarization-spatial collective unitary operation}\label{Sec2.1}

Based on Eqs. (\ref{eq6}-\ref{eq7}), arbitrary two-qubit polarization-spatial collective unitary operation $U_{\text{ps}}\in\text{SU}(4)$ group can be decomposed as
\begin{eqnarray}    \label{eq15}
\begin{split}
U_{\text{ps}}&= \text{S}[\text{U}(2)\oplus \text{U}(2)] \cdot A \cdot  \text{S}[\text{U}(2)\oplus \text{U}(2)] \\ &
\overset{\text{def}}{=} (L_1 \oplus L_2) \cdot A \cdot (R_1 \oplus R_2),
\end{split}
\end{eqnarray}
in the basis $\{|H a_1 \rangle$, $|H a_2\rangle$, $|V a_1\rangle$, $|V a_2\rangle\}$. Here $H$ and $V$ represent the horizontal and vertical polarization DOF of a single photon, respectively. $a_1$ and $a_2$ represent two spatial-mode DOF of a single photon. $L_1$ and $L_2$ ($R_1$ and $R_2$) are the local polarization single-qubit gates acting on each spatial-mode  separately. $A$ is created by the Cartan subalgebra $\mathfrak{h}_{2^2}$ with a block diagonal form
\begin{eqnarray}    \label{eq16}
\begin{split}
A&=e^{i(\alpha \mathbb{I}_2\otimes \sigma_x +\beta \sigma_z\otimes \sigma_x)}.
\end{split}
\end{eqnarray}
We denote $\theta_1=\alpha+\beta$ and $\theta_2=\alpha-\beta$. In order to realize the polarization-spatial operation $A$ efficiently, we further decompose $A$ as
%
\begin{eqnarray}    \label{eq17}
\begin{split}
A=&
\left(
     \begin{array}{cccc}
       0 & 0 & i  & 0 \\
       0 & 1 & 0  & 0 \\
       i & 0 & 0  & 0 \\
       0 & 0 & 0  & -1 \\
     \end{array}
\right)\cdot
\left(
     \begin{array}{cccc}
       0 & 1 & 0 & 0 \\
       1 & 0 & 0 & 0 \\
       0 & 0 & 1 & 0 \\
       0 & 0 & 0 & 1 \\
     \end{array}
\right) \\ &\cdot
\left(
     \begin{array}{cccc}
     i\cos\theta_1 & 0                  & i\sin\theta_1 & 0 \\
       0                & i\cos\theta_2 & 0                   &i\sin\theta_2 \\
     i\sin\theta_1& 0                  & -i\cos\theta_1 & 0                   \\
     0                  & i\sin\theta_2&  0                  &-i\cos\theta_2\\
     \end{array}
\right) \\&\cdot
\left(
     \begin{array}{cccc}
       0 & 1 & 0 & 0 \\
       1 & 0 & 0 & 0 \\
       0 & 0 & 1 & 0 \\
       0 & 0 & 0 & 1 \\
     \end{array}
\right)\cdot
\left(
     \begin{array}{cccc}
       0 & 0 & -1 & 0 \\
       0 & -i & 0 & 0 \\
       1 & 0 & 0  & 0 \\
       0 & 0 & 0  & -i \\
     \end{array}
\right).
\end{split}
\end{eqnarray}

In Eq. (\ref{eq17}), the second and the fourth matrices  represent a CNOT gate in the basis $\{|H a_1 \rangle$, $|H a_2\rangle$, $|V a_1\rangle$, $|V a_2\rangle\}$, which can be implemented conveniently by using a polarizing beam splitter (PBS), because the PBS transmits the $H$-polarized photon and reflects the $V$-polarized photon. The third matrix can be completed by two HWPs rotated to angles $\frac{\theta_1}{2}$ and $\frac{\theta_2}{2}$ acting on spatial modes $a_1$ and $a_2$, separately. The first and the last matrices can be absorbed into the single-qubit  operations $L_1$, $L_2$, $R_1$, and $R_2$, i.e., $i\sigma_x$ is absorbed into $L_1$, $\sigma_z$ is absorbed into $L_2$, $-i\sigma_y$ is absorbed into $R_1$, and $-i\mathbb{I}_2$ is absorbed into $R_2$.

\begin{figure} 
\begin{center}
\includegraphics[width=8 cm,angle=0]{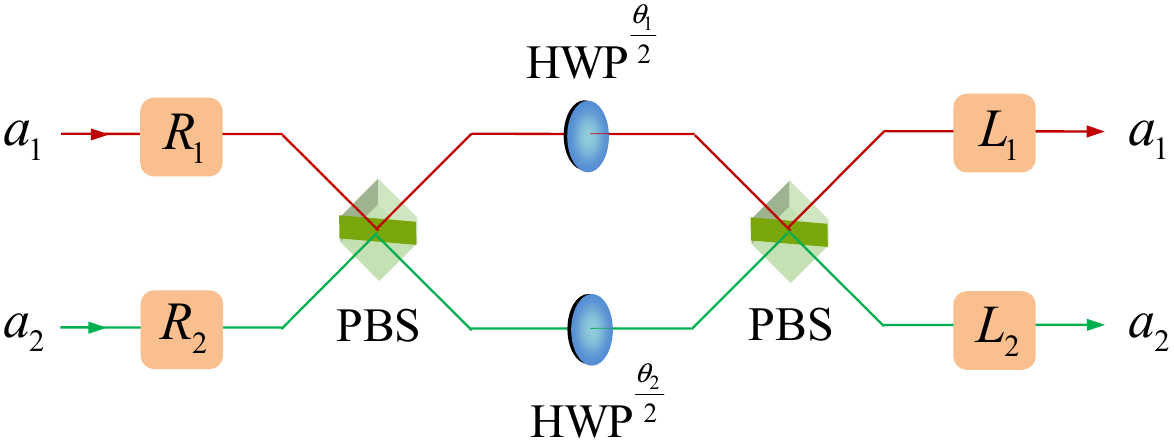}
\caption{Implementing arbitrary two-qubit polarization-spatial collective unitary operation with up to 20 linear optics. $a_1$ and $a_2$  represent two different spatial modes of a single photon. PBS is the polarizing beam splitter. HWP$^{\theta}$ denotes a half-wave plate rotated to angle $\theta$. $L_1$ and $R_1$ ($L_2$ and $R_2$) are polarization single-qubit gates acting on spatial-mode $a_1$ ($a_2$).} \label{polarization-cartan}
\end{center}
\end{figure}

Putting all the pieces together, as shown in Fig. \ref{polarization-cartan}, one can find that 20 linear optical elements are sufficient to realize an arbitrary two-qubit polarization-spatial collective unitary operation. Each of $L_1$, $L_2$, $R_1$, and $R_2$ in Fig. \ref{polarization-cartan}  can be efficiently achieved by employing at most two QWPs, one HWP and one PS \cite{single-qubit1,single-qubit2,single-qubit3}. The effects of optical elements PS, HWP, and QWP oriented to angle $\theta$  can be expressed by \cite{single-qubit1}
\begin{eqnarray}    \label{eq12}
\begin{split}
&U_{\text{PS}^\theta}=e^{i\theta}\mathbb{I}_2,\\
%
&U_{\text{HWP}^\theta}=e^{i\frac{\pi}{2}}\left(\begin{array}{cc}
\cos2\theta   &  \sin2\theta    \\
\sin2\theta    &  -\cos2\theta  \\
\end{array}\right),\\
%
%
&U_{\text{QWP}^\theta}=\frac{1}{\sqrt{2}}\left(\begin{array}{cc}
1+i\cos2\theta   &  i\sin2\theta    \\
i\sin2\theta   &  1-i\cos2\theta  \\
\end{array}\right).
\end{split}
\end{eqnarray}

\subsection{Application: the polarization-spatial quantum walk and quantum Fourier transformation}\label{Sec2.2}

Quantum walk and quantum Fourier transformation are both used as important benchmarks for realizing QIP tasks.
Two-dimensional quantum walk and quantum Fourier transformation matrices are described by \cite{Walk,book}
\begin{eqnarray}    \label{eq18}
U_{w}=\frac{1}{2}\left(\begin{array}{cccc}
-1   &  1   & 1  & 1\\
1    &  -1  & 1  & 1\\
1    &  1   &-1  & 1 \\
1    &  1   & 1  & -1\\
\end{array}\right),
\end{eqnarray}
\begin{eqnarray}    \label{eq19}
U_{F}=\frac{1}{2}\left(\begin{array}{cccc}
1   &  1   & 1  & 1\\
1   &  i   & -1 & -i\\
1   & -1   & 1  & -1 \\
1   & -i   & -1 & i\\
\end{array}\right).
\end{eqnarray}

\begin{figure} [htb]   
\begin{center}
\includegraphics[width=8 cm,angle=0]{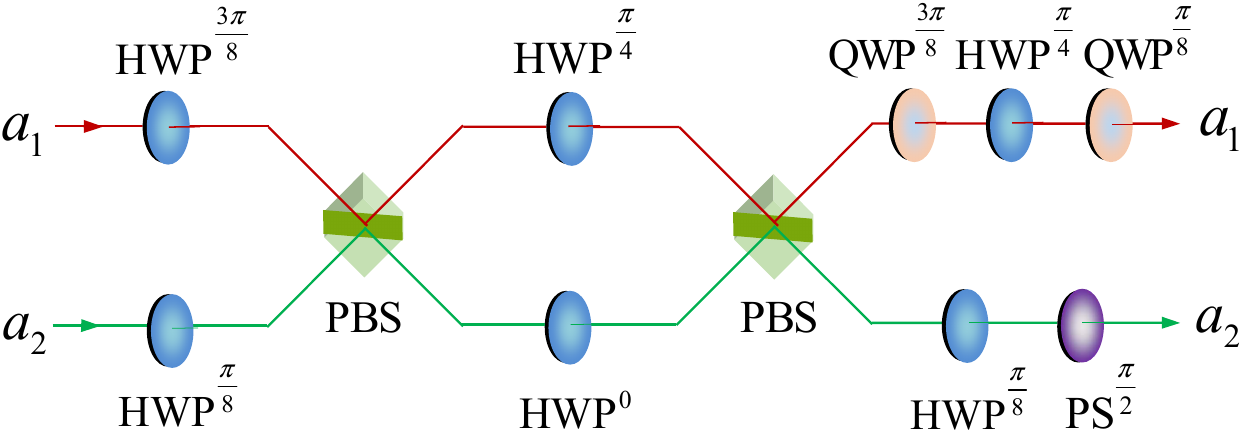}
\caption{ The scheme for implementing a two-dimensional polarization-spatial quantum walk.} \label{walk}
\end{center}
\end{figure}

Based on  Cartan decomposition described in Eqs. (\ref{eq6}-\ref{eq7}), $U_{w}$  can be decomposed as
\begin{eqnarray}
\begin{split}    \label{eq22}
U_{w}=&
\left(
  \begin{array}{cccc}
    -1 & 0  & -1 & 0 \\
    0  & -1 & 0  & -1 \\
    1  & 0  & -1 & 0 \\
    0  & -1 & 0  & 1 \\
  \end{array}
\right)\cdot
\left(
  \begin{array}{cccc}
    i & 0 & 0 & 0 \\
    0 & 0 & i & 0 \\
    0 & i & 0 & 0 \\
    0 & 0 & 0 & -i \\
  \end{array}
\right)   \\ &  \cdot\frac{i}{2}
\left(
  \begin{array}{cccc}
    -1 & 0   & 1  & 0 \\
    0  & 1   & 0  & 1 \\
    1  & 0   & 1  & 0 \\
    0  & 1   & 0  &-1 \\
  \end{array}
\right).
\end{split}
\end{eqnarray}
Here the second matrix  $A_w$ in Eq. (\ref{eq22}) can be further factorized as
\begin{eqnarray}    \label{eq23}
\begin{split}
A_w=&
\left(
     \begin{array}{cccc}
       0 & 1 & 0 & 0 \\
       1 & 0 & 0 & 0 \\
       0 & 0 & 1 & 0 \\
       0 & 0 & 0 & 1 \\
     \end{array}
\right)  \cdot
\left(
     \begin{array}{cccc}
     0 & 0  & i  & 0 \\
     0 & i  & 0  &0  \\
     i & 0  & 0  & 0  \\
     0 & 0  & 0  &-i \\
     \end{array}
\right)  \\&  \cdot
\left(
     \begin{array}{cccc}
       0 & 1 & 0 & 0 \\
       1 & 0 & 0 & 0 \\
       0 & 0 & 1 & 0 \\
       0 & 0 & 0 & 1 \\
     \end{array}
\right).
\end{split}
\end{eqnarray}
Based on Eqs. (\ref{eq22}-\ref{eq23}),  we design a quantum circuit to implement a deterministic two-dimensional quantum walk in the basis $\{|H a_1 \rangle$, $|H a_2\rangle$, $|V a_1\rangle$, $|V a_2\rangle\}$ with 11 linear optics, see Fig. \ref{walk}.

\begin{figure} [htb]   
\begin{center}
\includegraphics[width=8 cm,angle=0]{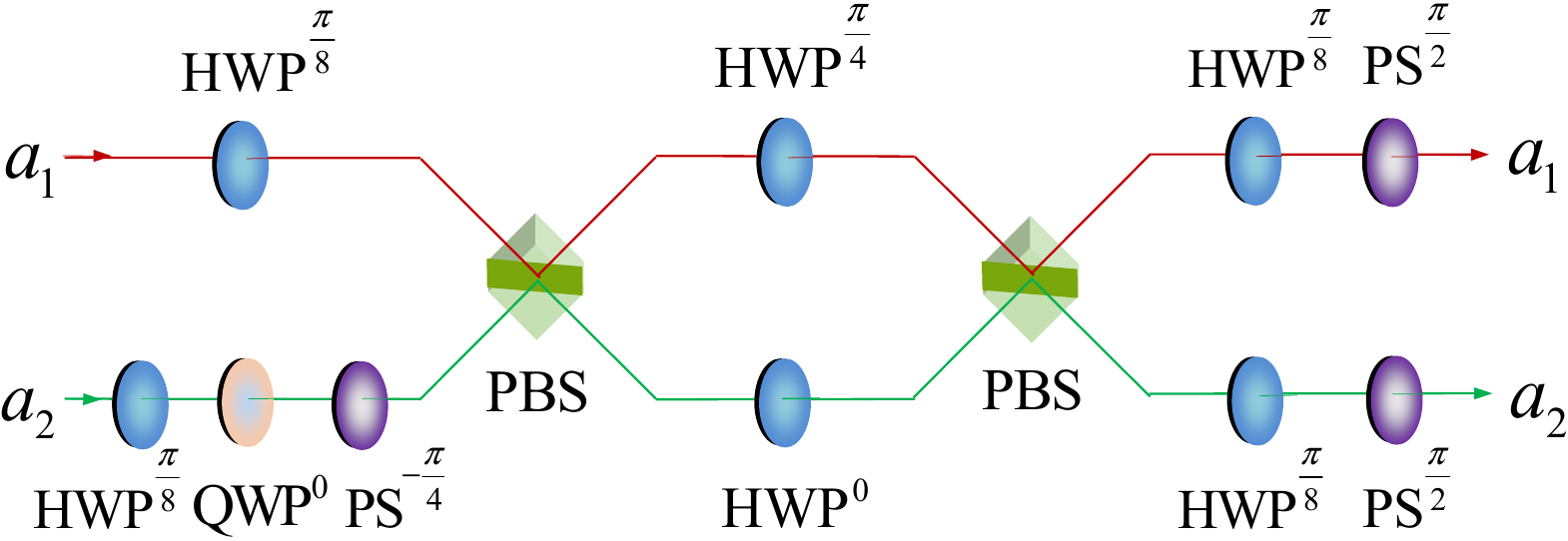}
\caption{ The scheme for implementing a two-qubit polarization-spatial quantum Fourier transformation.} \label{Fourier}
\end{center}
\end{figure}

Similarly, two-qubit quantum Fourier transformation  can be decomposed as
\begin{eqnarray}
\begin{split}     \label{eq26}
U_{F}=&
 \left(
  \begin{array}{cccc}
    -1 & 0  & -1 & 0 \\
    0  & -1 & 0  & -1 \\
    -1 & 0  & 1  & 0 \\
    0  & -1 & 0  & 1 \\
  \end{array}
\right)  \cdot
\left(
  \begin{array}{cccc}
     i & 0 & 0 & 0 \\
    0 & 0 & i & 0 \\
    0 & i & 0 & 0 \\
    0 & 0 & 0 & -i \\
  \end{array}
\right)   \\ &  \cdot \frac{i}{2}
\left(
  \begin{array}{cccc}
    1  & 0  & 1   & 0 \\
    0  & 1  & 0   & 1 \\
    1  & 0  & -1  & 0 \\
    0  & -i & 0   &i \\
  \end{array}
\right).
\end{split}
\end{eqnarray}
Based on Eq. (\ref{eq26}), Fig. \ref{Fourier} shows a specific setup for implementing  two-qubit polarization-spatial quantum Fourier transform with 12 linear optics.

\section{Arbitrary spatial-polarization collective unitary operation}\label{Sec3}

We note that Cartan decomposition is not unique, and the form of decomposition depends on the span  of the Lie subalgebra and the Cartan subalgebra. We next investigate the decomposition for arbitrary spatial-polarization  collective unitary operation.

For $n=1$, the Cartan decomposition of the Lie algebra $\mathfrak{su}(2)$ can still be taken the form
\begin{eqnarray}    \label{eq28}
\begin{split}
\tilde{\mathfrak{l}}_{2^1}:=\text{span}\{\sigma_z\}, \quad \tilde{\mathfrak{p}}_{2^1}:=\text{span}\{\sigma_x, \sigma_y \}.
\end{split}
\end{eqnarray}
The Cartan subalgebra $\tilde{\mathfrak{h}}_{2^1}$ is taken as
\begin{eqnarray}    \label{eq29}
\begin{split}
\tilde{\mathfrak{h}}_{2^1}&:=\text{span}\{\sigma_x\}.
\end{split}
\end{eqnarray}

For $n=2$, in order to realize the spatial-polarization collective unitary operation,  $\mathfrak{su}(4)$ can be factorized as %
\begin{eqnarray}    \label{eq30}
\begin{split}
\tilde{\mathfrak{l}}_{2^2}&:=\text{span}\{\mathbb{I}_2 \otimes \mathfrak{su}(2), \sigma_z \otimes \mathfrak{u}(2)\},\\
\tilde{\mathfrak{p}}_{2^2}&:=\text{span}\{\sigma_x \otimes \mathfrak{u}(2), \sigma_y \otimes \mathfrak{u}(2) \},
\end{split}
\end{eqnarray}
with Cartan subalgebra 
\begin{eqnarray}    \label{eq31}
\begin{split}
\tilde{\mathfrak{h}}_{2^2}&:=\text{span}\{\sigma_x\otimes \sigma_y, \sigma_y\otimes \sigma_x \}.
\end{split}
\end{eqnarray}

\begin{figure}     [htb]
\begin{center}
\includegraphics[width=8 cm,angle=0]{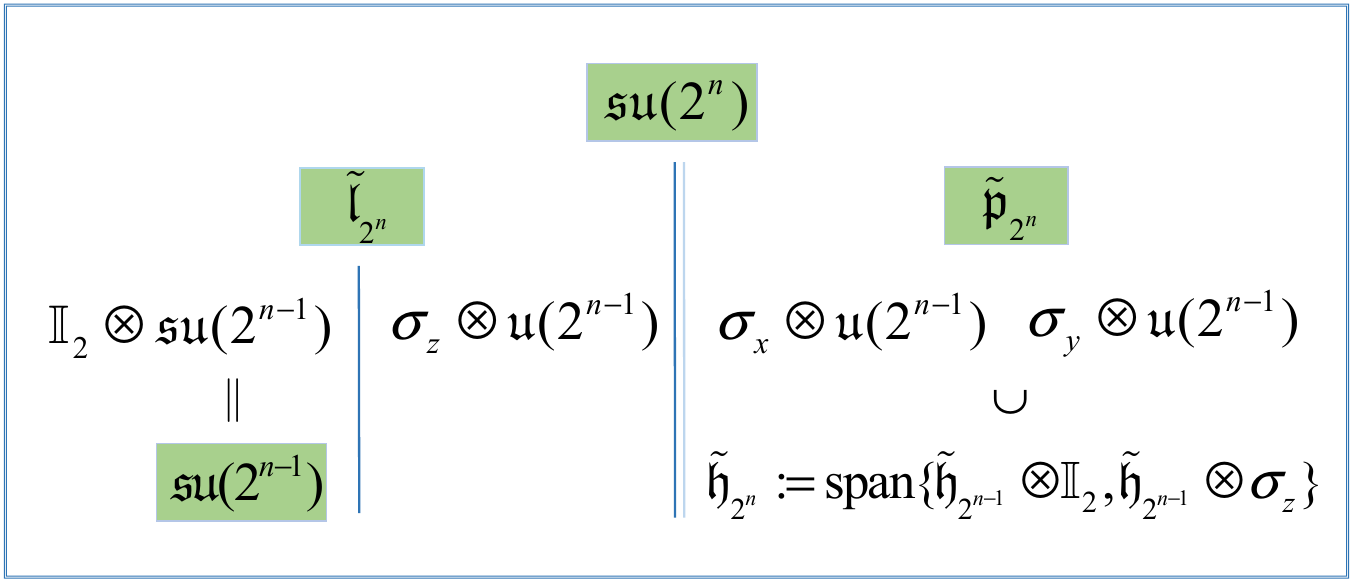}
\caption{ Schematic explanation for Cartan decomposition of a spatial-polarization collective unitary operation by a recursive algorithm. }  \label{SP-su(2n)}
\end{center}
\end{figure}

For an $n$-qubit ($n>2$) case, as  shown in Fig.  \ref{SP-su(2n)}, the factorization of the $\mathfrak{su}(2^n)$ can be expressed by
\begin{eqnarray}    \label{eq32}
\begin{split}
\tilde{\mathfrak{l}}_{2^n}&:=\text{span}\{\mathbb{I}_2 \otimes \mathfrak{su}(2^{n-1}), \sigma_z \otimes \mathfrak{u}(2^{n-1})\},\\
\tilde{\mathfrak{p}}_{2^n}&:=\text{span}\{ \sigma_x \otimes \mathfrak{u}(2^{n-1}),  \sigma_y \otimes \mathfrak{u}(2^{n-1})\},
\end{split}
\end{eqnarray}
with Cartan subalgebra
\begin{eqnarray}    \label{eq33}
\begin{split}
\tilde{\mathfrak{h}}_{2^n}&:=\text{span}\{\tilde{\mathfrak{h}}_{2^{n-1}}  \otimes  \mathbb{I}_2,   \tilde{\mathfrak{h}}_{2^{n-1}} \otimes  \sigma_z\}.
\end{split}
\end{eqnarray}

\subsection{Implementation of arbitrary two-qubit spatial-polarization collective unitary operation}\label{Sec3.1}

\begin{figure} 
\begin{center}
\includegraphics[width=8  cm,angle=0]{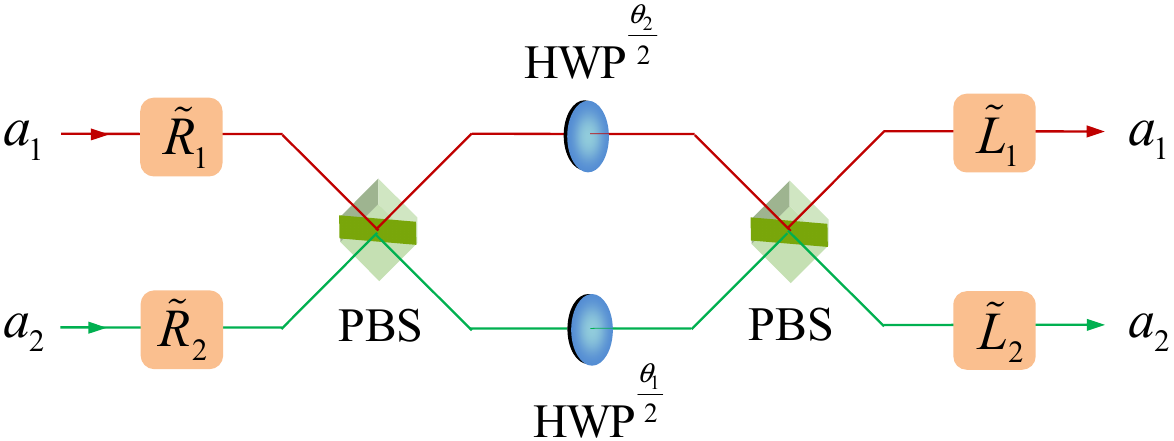}
\caption{Implementing arbitrary two-qubit spatial-polarization collective unitary operation with up to 20 linear optics.} \label{SP-cartan}
\end{center}
\end{figure}

Based on Eqs. (\ref{eq30}-\ref{eq31}), one can see that arbitrary two-qubit collective unitary operation $U_{\text{sp}}\in\text{SU}(4)$ group in the basis $\{|a_1 H\rangle$, $|a_1 V \rangle$, $| a_2 H\rangle$,  $ |a_2 V\rangle\}$ can be factorized as
\begin{eqnarray} \label{eq34}
\begin{split}
U_{\text{sp}}&=\text{S}[\text{U}(2)\oplus \text{U}(2)]\cdot \widetilde{A}\cdot \text{S}[\text{U}(2)\oplus \text{U}(2)]\\  &
\overset{\text{def}}{=}\left(
    \begin{array}{cc}
      \widetilde{L}_1 & 0 \\
      0 & \widetilde{L}_2 \\
    \end{array}
  \right)
\cdot \widetilde{A}\cdot
\left(
    \begin{array}{cc}
      \widetilde{R}_1 & 0 \\
      0 & \widetilde{R}_2 \\
    \end{array}
  \right).
\end{split}
\end{eqnarray}
Here $\widetilde{L}_1$ ($\widetilde{L}_2$) and $\widetilde{R}_1$ ($\widetilde{R}_2$) are arbitrary single-qubit polarization gates acting on the spatial-mode $a_1$ ($a_2$). $\widetilde{A}$ can be generated by $\tilde{\mathfrak{h}}_{2^2}$ with the following form
\begin{eqnarray}    \label{eq35}
\begin{split}
\widetilde{A}&=e^{i(\alpha \sigma_x\otimes \sigma_y +\beta \sigma_y\otimes \sigma_x)}.
\end{split}
\end{eqnarray}
$\widetilde{A}$ can be further factorized  as
\begin{eqnarray}    \label{eq36}
\begin{split}
\widetilde{A}=&
\left(
     \begin{array}{cccc}
      -i & 0 & 0  & 0 \\
       0 & i & 0  & 0 \\
       0 & 0 & -i & 0 \\
       0 & 0 & 0  & i \\
     \end{array}
\right)\cdot
\left(
     \begin{array}{cccc}
       0 & 0 & 1 & 0 \\
       0 & 1 & 0 & 0 \\
       1 & 0 & 0 & 0 \\
       0 & 0 & 0 & 1 \\
     \end{array}
\right)\\ & \cdot
\left(
     \begin{array}{cccc}
       i\cos \theta_2 & i\sin \theta_2  & 0 & 0 \\
       i\sin \theta_2  & -i\cos \theta_2  & 0 & 0 \\
       0 & 0 & i\cos \theta_1  & i\sin \theta_1  \\
       0 & 0 & i\sin \theta_1  & -i\cos \theta_1  \\
     \end{array}
\right)\\&\cdot
\left(
     \begin{array}{cccc}
      0 & 0 & 1 & 0 \\
       0 & 1 & 0 & 0 \\
       1 & 0 & 0 & 0 \\
       0 & 0 & 0 & 1 \\
     \end{array}
\right).
\end{split}
\end{eqnarray}
The second  and the fourth matrices in Eq. (\ref{eq36}) denote a spatial-polarization CNOT gate which can be easily realized by a PBS. The third matrix can be achieved by two HWPs with the angles $\frac{\theta_2}{2}$ and $\frac{\theta_1}{2}$ acting on spatial-modes $a_1$ and $a_2$, respectively. The first matrix will be absorbed into $\widetilde{L}_1$ and $\widetilde{L}_2$. Each of single-qubit operations $\widetilde{L}_1$, $\widetilde{L}_2$, $\widetilde{R}_1$, and $\widetilde{R}_2$, can be completed by at most two QWPs, one HWP and one PS \cite{single-qubit1,single-qubit2,single-qubit3}.  Figure  \ref{SP-cartan} presents the setup to implement arbitrary two-qubit spatial-polarization collective unitary operation with at most 20 linear optics.

\subsection{Application: the spatial-polarization quantum walk and quantum Fourier transformation}\label{Sec3.2}

\begin{figure} [htb]   
\begin{center}
\includegraphics[width=8 cm,angle=0]{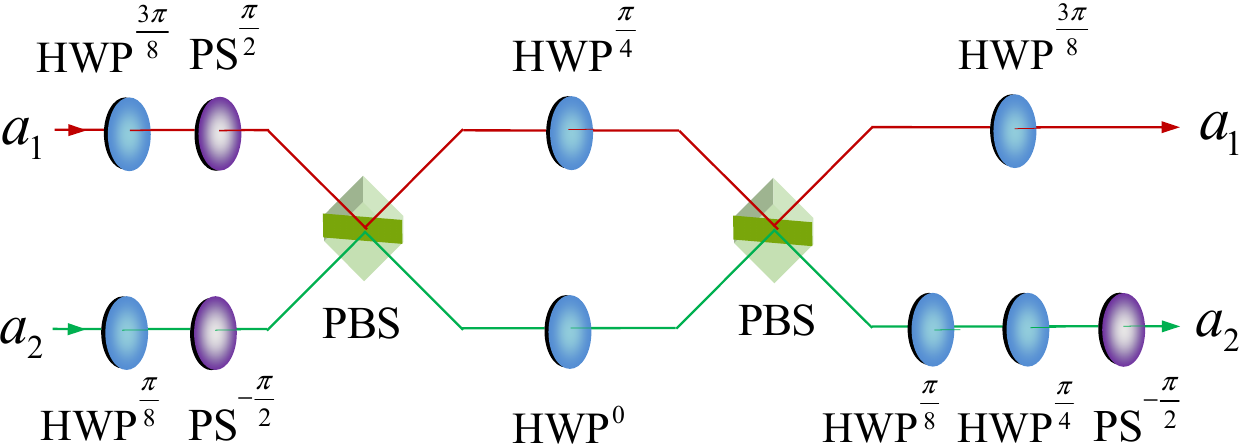}
\caption{ The scheme for implementing a two-dimensional spatial-polarization quantum walk.} \label{SPwalk}
\end{center}
\end{figure}

In order to realize spatial-polarization quantum walk, based on Eqs. (\ref{eq30}-\ref{eq31}), two-dimensional quantum walk can be redecomposed as
\begin{eqnarray}
\begin{split}    \label{eq39}
U_{w}=&
\left(
  \begin{array}{cccc}
    -1 & -1 & 0& 0 \\
    1 & -1  & 0 & 0 \\
    0 & 0 & -1 & -1 \\
    0 & 0 & -1 & 1 \\
  \end{array}
\right)  \cdot
\left(
  \begin{array}{cccc}
    1 & 0 & 0 & 0 \\
    0 & 0 & -1 & 0 \\
    0 & 1 & 0 & 0 \\
    0 & 0 & 0 & 1 \\
  \end{array}
\right) \\ & \cdot\frac{1}{2}
\left(
  \begin{array}{cccc}
    1 & -1 & 0& 0 \\
    -1 & -1  & 0 & 0 \\
    0 & 0 & 1 & 1 \\
    0 & 0 & 1 & -1 \\
  \end{array}
\right).
\end{split}
\end{eqnarray}
The second matrix $\widetilde{A}_w$ in Eq. (\ref{eq39}) can be further decomposed as
\begin{eqnarray}    \label{eq40}
\begin{split}
\widetilde{A}_w=&
\left(
     \begin{array}{cccc}
       -i & 0 & 0 & 0 \\
       0 & i & 0 & 0 \\
       0 & 0 & -i & 0 \\
       0 & 0 & 0 & i \\
     \end{array}
\right)\cdot
\left(
     \begin{array}{cccc}
       0 & 0 & 1 & 0 \\
       0 & 1 & 0 & 0 \\
       1 & 0 & 0 & 0 \\
       0 & 0 & 0 & 1 \\
     \end{array}
\right) \\& \cdot
\left(
     \begin{array}{cccc}
     0 & i  & 0  & 0 \\
     i & 0  & 0  &0  \\
     0 & 0  & i  & 0  \\
     0 & 0  & 0  &-i \\
     \end{array}
\right)  \cdot
\left(
     \begin{array}{cccc}
       0 & 0 & 1 & 0 \\
       0 & 1 & 0 & 0 \\
       1 & 0 & 0 & 0 \\
       0 & 0 & 0 & 1 \\
     \end{array}
\right).
\end{split}
\end{eqnarray}
Based on Eqs. (\ref{eq39}-\ref{eq40}), Fig. \ref{SPwalk} shows the setup to implement a two-dimensional spatial-polarization quantum walk with 12 optical elements.

\begin{figure} [!h]   
\begin{center}
\includegraphics[width=8.3 cm,angle=0]{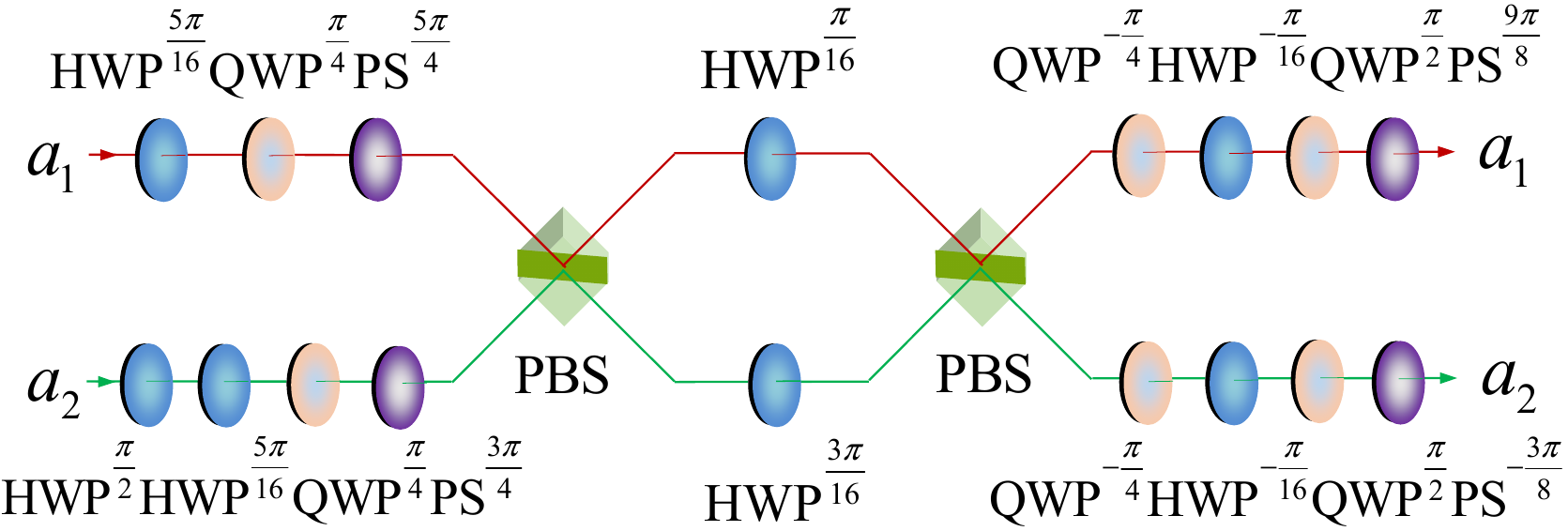}
\caption{The scheme for implementing a two-qubit spatial-polarization quantum Fourier transformation.} \label{SP-Fourier}
\end{center}
\end{figure}

Similarly, we redecompose two-qubit quantum Fourier transformation matrix in  Eq. (\ref{eq41}).
\begin{eqnarray}    \label{eq41}
\begin{split}
U_{F}=&
\text{diag} \bigg \{\left(
  \begin{array}{cc}
    i\sin\frac{\pi}{8}-\cos\frac{\pi}{8} & -\sin\frac{\pi}{8}-i\cos\frac{\pi}{8}  \\
    \cos\frac{\pi}{8}+i\sin\frac{\pi}{8}  & \sin\frac{\pi}{8}-i\cos\frac{\pi}{8}  \\
      \end{array}\right), \\&
  \left(
  \begin{array}{cc}
     \cos\frac{\pi}{8}-i\sin\frac{\pi}{8}  & -\sin\frac{\pi}{8}-i\cos\frac{\pi}{8} \\
     -\cos\frac{\pi}{8}-i\sin\frac{\pi}{8}  & \sin\frac{\pi}{8}-i\cos\frac{\pi}{8} \\
  \end{array}\right)\bigg\}
  \\ &\cdot
\left(
  \begin{array}{cccc}
    \cos\frac{3\pi}{8} & 0 & 0 & \sin\frac{3\pi}{8} \\
    0 & \cos\frac{\pi}{8} & -\sin\frac{\pi}{8} & 0 \\
    0 & \sin\frac{\pi}{8} & \cos\frac{\pi}{8} & 0 \\
    -\sin\frac{3\pi}{8} & 0 & 0 & \cos\frac{3\pi}{8} \\
  \end{array}
\right) \\ &\cdot\frac{1}{2}
\left(
  \begin{array}{cccc}
    -i  &  \frac{i-1}{\sqrt{2}}  & 0  & 0 \\
     i  &  \frac{i-1}{\sqrt{2}}  & 0  & 0 \\
     0  & 0   & 1 & -\frac{i+1}{\sqrt{2}} \\
     0  & 0   & -1  & -\frac{i+1}{\sqrt{2}} \\
  \end{array}
\right).
\end{split}
\end{eqnarray}
The second matrix $\widetilde{A}_F$ in Eq. (\ref{eq41}) can be further factorized as
\begin{eqnarray}    \label{eq42}
\begin{split}
\widetilde{A}_F=&
\left(
     \begin{array}{cccc}
       -i & 0 & 0 & 0 \\
       0 & i & 0 & 0 \\
       0 & 0 & -i & 0 \\
       0 & 0 & 0 & i \\
     \end{array}
\right)\cdot
\left(
     \begin{array}{cccc}
       0 & 0 & 1 & 0 \\
       0 & 1 & 0 & 0 \\
       1 & 0 & 0 & 0 \\
       0 & 0 & 0 & 1 \\
     \end{array}
\right) \\& \cdot
\left(
     \begin{array}{cccc}
     i\cos\frac{\pi}{8} & i\sin\frac{\pi}{8}  & 0    & 0 \\
     i\sin\frac{\pi}{8} & -i\cos\frac{\pi}{8} & 0   &0  \\
     0 & 0    & i\cos\frac{3\pi}{8} & i\sin\frac{3\pi}{8}  \\
     0 & 0    & i\sin\frac{3\pi}{8} & -i\cos\frac{3\pi}{8}  \\
     \end{array}
\right) \\&  \cdot
\left(
     \begin{array}{cccc}
       0 & 0 & 1 & 0 \\
       0 & 1 & 0 & 0 \\
       1 & 0 & 0 & 0 \\
       0 & 0 & 0 & 1 \\
     \end{array}
\right).
\end{split}
\end{eqnarray}
Figure \ref{SP-Fourier} shows the setup to implement a two-qubit spatial-polarization quantum Fourier transformation with 19 linear optical elements.

\begin{figure} [htb]    
\begin{center}
\includegraphics[width=8.1 cm,angle=0]{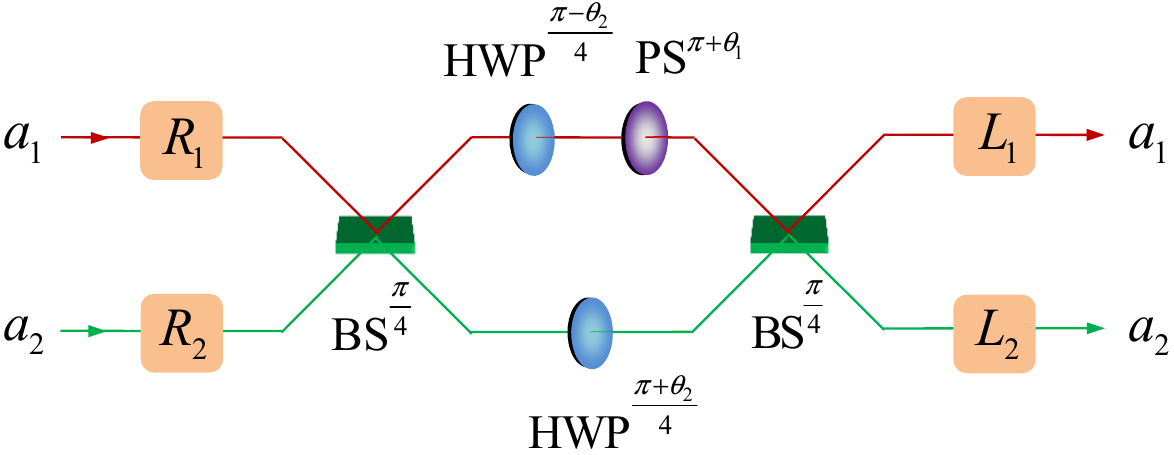}
\caption{CSD-based arbitrary two-qubit spatial-polarization collective unitary operation \cite{Dhand-Goyal}.  Beam splitter BS$^{\frac{\pi}{4}}$ completes operations $|a_1\rangle\leftrightarrow (|a_1\rangle-|a_2\rangle)/\sqrt{2}$ and $|a_2\rangle\leftrightarrow (|a_1\rangle+|a_2\rangle)/\sqrt{2}$. }   \label{spatial-CSD}
\end{center}
\end{figure}

\begin{figure} [htb]    
\begin{center}
\includegraphics[width=8.5 cm,angle=0]{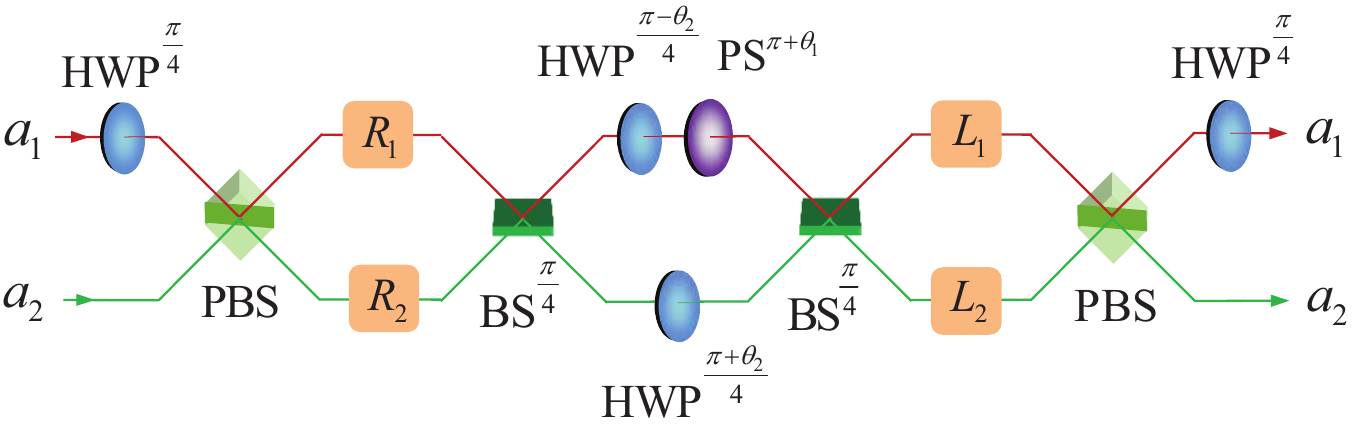}
\caption{CSD-SWAP-based arbitrary two-qubit polarization-spatial collective unitary operation \cite{Dhand-Goyal}.} \label{polarization-CSD}
\end{center}
\end{figure}

\begin{figure} [!h]     
\begin{center}
\includegraphics[width=8.6 cm,angle=0]{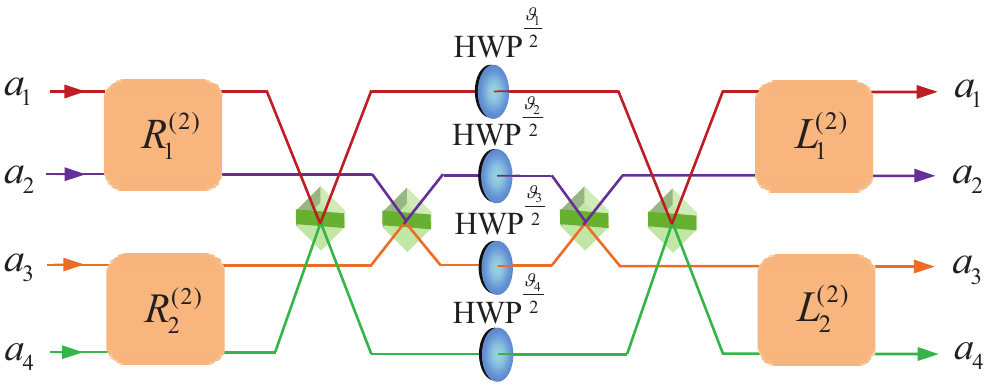}
\caption{ Implementation of arbitrary unitary operation with spatial modes $m$=4 and polarization modes $n=2$. $L^{(2)}_1$ ($L^{(2)}_2$) and $R^{(2)}_1$ ($R^{(2)}_2$) are two-qubit gates acting on spatial-modes $a_1$ and $a_2$ ($a_3$ and $a_4$), which can be realized by Fig. \ref{SP-cartan}.} \label{m=4sp}
\end{center}
\end{figure}

\section{Conclusion} \label{Sec4}

In this letter, by devising Cartan decompositions in detail, we have presented two deterministic optical architectures to implement arbitrary two-qubit collective unitary operation in polarization-spatial DOF and spatial-polarization DOF of a single photon, respectively.  The parameterized quantum computation can be flexibly controlled by adjusting the angles of HWP, QWP, and PS. The number of required linear optical elements for two-qubit spatial-polarization scheme is reduced from 21 \cite{Dhand-Goyal} (the CSD-based circuit, see Fig. \ref{spatial-CSD}) to 20, and the polarization-spatial one is reduced from 25 \cite{Dhand-Goyal} (the CSD-SWAP-based circuit, see Fig. \ref{polarization-CSD}) to 20. Our method is also suitable for more spatial modes ($m>2$). For example, based on the iterative spatial-polarization Cartan decomposition in Fig. \ref{SP-su(2n)}, arbitrary collective unitary operation with spatial modes $m$=4  and   polarization modes $n=2$ can be realized with 48 linear optical elements (see Fig. \ref{m=4sp}), which improves on the CSD-based one with 74 optical elements \cite{Dhand-Goyal}.

As an application, we designed four compact schemes to implement two-dimensional quantum walk and quantum Fourier transformation. Our perspective may have potential advantages in simplifying quantum computing and quantum communication using multiple DOFs encoding.

\section*{Acknowledgements} \par
This work is supported by the National Natural Science Foundation of China under Grant No. 11604012,  the Fundamental Research Funds for the Central Universities under Grant No. FRF-TP-19-011A3, and a grant from the China Scholarship Council.

\end{document}